\documentclass[article,showpacs]{revtex4}
\usepackage{graphicx}
\begin{document}
\title{Bound states in the phase diagram of the extended Hubbard model}
\author{Maciej Bak}
\email[]{karen@delta.amu.edu.pl}
\affiliation{Institute of
Physics, A. Mickiewicz University, Umultowska 85, 61-614 Pozna\'n,
Poland}

\begin{abstract}
The paper shows how the known, exact results for the two electron
bound states can modify the ground state phase diagram of extended
Hubbard model (EHM) for on-site attraction, intersite repulsion
and arbitrary electron density. The main result is suppression of
the superconducting state in favor of normal phase  for small
charge densities.
\end{abstract}


\maketitle

\section{Introduction}
The Hubbard model appears in almost all areas of solid state
physics. Its universality is connected with the fact that it
describes both band movement of charges as well as local and
nonlocal -- in the extended model -- correlations~\cite{hubbard}.
Treating its parameters as effective ones, the model has been used
in research of magnetism, superconductivity and especially high
temperature superconductivity (HTS), other various phenomena in
the solid, charge orderings, phase separation etc., in materials
like high temperature superconductors, bismuthates, Chevrel
phases, amorphous semiconductors, heavy fermion materials, systems
with alternating valence to name the few (for a review see, e.g.,
Ref.~\cite{review}).

Unfortunately there are not many exact results concerning this
model. Usually the results are obtained in specific limits:
infinite dimensions, one dimension (the most numerous group),
infinite repulsion, half-filled band or other specific band
fillings. We can mention the exact solution in one dimension
($d=1$) obtained by Bethe ansatz~\cite{betheA}, Lieb's
ferrimagnetism~\cite{lieb}, Nagaoka ferromagnetism in repulsive,
half-filled systems with one hole~\cite{nagaoka}, flat band
ferromagnetism~\cite{mielke}, some bounds on correlation
functions~\cite{koma,kubo,shen} and a finding of
Randeria~\cite{randeria}, according to which, existence of
two-electron bound states of s-wave symmetry is necessary and
sufficient condition for appearance of s-wave superconductivity in
 $d=2$ systems with low electron density. Let's also note that
the mean-field BCS equations for superconductivity in the Hubbard
model with effective attractive interaction between electrons, in
the limit of vanishing electron density turn into Schrodinger
equations, which can also be solved exactly~\cite{nozieres}.

A phase diagram in two dimensions for arbitrary $n$, a case of
special interest due to the possible connection with high
temperature superconductivity, is still under examination. The
results obtained in the mean-field approximation show competition
of phases: in half-filled band charge density waves (CDW) for
$W\ge 0$, superconductivity for $U<0$ and spin density waves for
$U>0$~\cite{review}. The calculations for $n\neq 1$ suggest
possibility of phase separation for $U<0$~\cite{rp,review}:
electron droplets for large enough $W<0$ and phase separation of
CDW (with $n=1$) with SS for $W>0$ (PS[CDW/SS]), for $n$ around
half-filled band competing with the pure SS state in low density
limit.

This paper shows, how the known solutions for the bound states
(including exact solutions of Schrodinger equation) can be used
for modification of the ground state phase diagram of the extended
Hubbard model.

\section{Hamiltonian and the superconducting state}
We begin with the extended Hubbard Hamiltonian in standard
notation:
\begin{equation}\label{ham}
    H=\sum_{ij\sigma}t_{ij}c^\dagger_{i\sigma}c_{j\sigma}+
        U \sum_i n_{i\uparrow}n_{i\downarrow}+
        \frac{1}{2}W \sum_{ij}n_{i\sigma}n_{j\sigma'}-
        \mu \sum_i n_i \; ,
\end{equation}
where we sum over nearest-neighbors (nn) sites only. $U$ and $W$
are treated as effective parameters. We use broken-symmetry
Hartree-Fock approach (for details see Ref.~\cite{mrrt}). As we
are interested in the properties of the superconducting state, we
introduce averages of operators $c_{-k\downarrow}c_{k\uparrow}$ in
Wick's-type decoupling~\cite{mahan} of the four-operator terms in
the Hamiltonian. Non-zero average of such pair-creating operators
means phase-coherence among pairs, i.e. a presence of
superconducting state, and serves as an order parameter (see
Eq.~(\ref{dq})).
\begin{equation}\label{ham2}
    H_0=\sum_{k\sigma}(\varepsilon_k-\overline\mu)c^\dagger_{k\sigma}c_{k\sigma}+
        \sum_k (\Delta_k c^\dagger_{k\uparrow}c^\dagger_{-k\downarrow}+h.c.)+C \; ,
\end{equation}
where:
\begin{equation}\label{dq}
    \Delta_{k_1}=\frac{1}{N}\sum_{k_2}(W_{k_2-k_1}+U)\langle c_{-k_2\downarrow}c_{k_2\uparrow}\rangle \; ,
\end{equation}
and $\overline\mu=\mu-(\frac{U}{2}+z W)n$, where $z$ is
coordination number of hypercubic lattice, $W_k=W\gamma_k$,
$\varepsilon_k=-t\gamma_k$, $\gamma_k=2\sum_{\alpha}^{d}\cos
k_\alpha$, $\alpha\in(x,y,z)$. After diagonalization of the
Hamiltonian Eq.~(\ref{ham2}) we obtain quasiparticle energy:
\begin{equation}\label{eq}
    E_q=\sqrt{(\varepsilon_q-\overline\mu\ )^2+|\Delta_q|^2} \; .
\end{equation}
and a self-consistent equation for the gap:
\begin{equation}
    \Delta_k=\frac{1}{N}\sum_q(W_{k-q}+U)\frac{\Delta_q}{2E_q}\tanh\frac{\beta E_q}{2} \; ,
\end{equation}
where $\beta=1/k_BT$, $T$ is temperature and $k_B$ Boltzmann
constant. The constant $C$ in the Hamiltonian Eq.~(\ref{ham2}) can
be expressed now as:
\begin{equation}
    C=-\frac{1}{4}(U+2W z)n^2+\frac{1}{N}\sum_k\frac{|\Delta_k|^2}{2E_k}\tanh\frac{\beta E_k}{2} \; ,\\
\end{equation}
The pairing potential in the singlet channel (see Eq.~(\ref{dq}))
 takes on the separable form for the square lattice and nn
interaction: $U+W_{k_1-k_2}=U+W\gamma_{k_1}\gamma_{k_2}/z$
(retaining only the terms of s-wave symmetry), and that makes
possible solving Eq.~(\ref{dq}) by an ansatz:
\begin{equation}\label{ansatz}
    \Delta_k=\Delta_0+\Delta_{\gamma} \gamma_k \; ,
\end{equation}
what leads us to the set of self-consistent equations:
\begin{eqnarray}\label{delta0}
    \Delta_0&=&-U \frac{1}{N}
        \sum_q(\Delta_0+\gamma_q\Delta_{\gamma})F_q\;,\\\label{delta1}%\frac{1}{2E_q}\tanh\frac{\beta E_q}{2} \; ,\\\label{delta1}
    \Delta_{\gamma}&=&-\frac{W}{z} \frac{1}{N}
        \sum_q\gamma_q(\Delta_0+\gamma_q\Delta_{\gamma})F_q\;,\\\label{n}%\frac{1}{2E_q}\tanh\frac{\beta E_q}{2} \; ,\\\label{n}
    n-1&=&-\frac{2}{N}\sum_q(\varepsilon_q-\overline\mu)F_q\;.%{2E_q}\tanh\frac{\beta E_q}{2} \; ,%\\
%    p&=&-\frac{1}{N}\sum_k\frac{{\tilde\varepsilon}_k\gamma_k}{2E_k}\tanh\frac{\beta E_q}{2} \; ,
\end{eqnarray}
where $F_q=(\tanh\frac{\beta E_q}{2})/2E_q$. In the case of the
rectangular density of states (DOS) and pure on-site pairing we
can obtain analytical solutions~\cite{rp}; in the case of the
extended s-wave superconductivity (Eqs~(\ref{delta0}) --
(\ref{n})) analytical solutions exist in the limit of low electron
density. Introducing a new parameter: $\Delta_\gamma/\Delta_0$, we
can expand Eqs~(\ref{delta0}) -- (\ref{n}), treating $\Delta_0$ as
a small parameter. As a result we  obtain a formula for critical
value for appearance of superconductivity for given $U$ and $n$ in
the ground state~\cite{bakmix}:
\begin{equation}\label{wcrmy}
    W_{cr}=\frac{8t^2}{\mu(1-n)+8tI-2\mu^2/U} \; ,\mbox{\hspace{1cm}where\hspace{1cm}}
    I=\int_{-1}^{\mu/D}x\rho(x)\;dx \; ,
\end{equation}
and $D=zt$ is half-bandwidth unit. In the case of rectangular DOS
this formula reduces to~\cite{bak2sol}:
\begin{equation}\label{wcrmysq}
    W_{cr}(1+(n-1)^2(1+16t/U))=-4t \; .
\end{equation}
We can go a step further in our mean-field analysis and include
Fock term $ p=\frac{1}{N}\sum_{k\sigma}\gamma_k\langle
c^\dagger_{k\sigma}c_{k\sigma}\rangle$ into calculations. In
Eqs~(\ref{ham2}), (\ref{eq}), (\ref{delta0}) -- (\ref{n}),
$\varepsilon_k$ must be changed into
${\tilde\varepsilon}_k=\varepsilon_k(1+pW/zt)$ then, and we have
to solve Eqs~(\ref{delta0}) -- (\ref{n}) self consistently with
the equation for the Fock term:
$p=-\frac{1}{N}\sum_k{\tilde\varepsilon}_k\gamma_k F_k$. Equations
(\ref{wcrmy}) -- (\ref{wcrmysq}) remain valid, with the change
$X\rightarrow X/(1+pW/zt)$ where $X=\mu$, $W_{cr}$ and $U$.

\section{Low density limit} Going back to Hamiltonian
Eq.~(\ref{ham}) we can obtain exact results in the low density
limit. In the center-of-mass coordinate system we can expand the
wave function of the two-electron bound pair $\psi$ in the basis
of plane waves (i.e., eigenstates of the hopping part of the
Hamiltonian Eq.~(\ref{ham})). We can easily find the equations for
the coefficients of the expansion, what finally yields  a set of
self-consistent equations for the wave function in the position
space, in terms of lattice Green functions~\cite{blaer,review}:
\begin{equation}\label{psiSelf}
    \psi(\textbf{r})=\sum_{\textbf{r}'}G(E,\textbf{P},\textbf{r},\textbf{r}')g(\textbf{r}')\psi(\textbf{r}') \; ,
\end{equation}
where $G$ is lattice Green function defined by:
\begin{equation}
    G(E,\textbf{P},\textbf{r},\textbf{r}')=\frac{1}{N}\sum_q\frac{\;e^{i\textbf{q}\cdot \textbf{r}}
        e^{-i\textbf{q}\cdot \textbf{r}'}}{E-E_{\textbf{Pq}}} \; ,
\end{equation}
and $g(\textbf{r})$ is diagonal interaction matrix, consisting of
elements $U$ and $W$. Eigenenergy equation takes the form:
\begin{equation}\label{det}
    det[{\cal G}-g^{-1}]=0 \; ,
\end{equation}
where $\cal{G}$ is a matrix with elements ${\cal
G}_{ij}=G(E,\textbf{P},\textbf{r}_i,\textbf{r}_j)$. This is an
analogue of Eqs~(\ref{delta0}) -- (\ref{delta1}). Let's note that
in the case of the two-electron bound pairs $\Delta=0$ and the
role of the binding energy is played by $\overline\mu/2$. For the
hypercubic lattices these equations were solved and it has been
found out that in one and two dimensions pairs for $W=0$ bind for
any negative $U$, while in three dimensions there is critical
value for $W$~\cite{mattis}. The formula for $W_{cr}$ in the case
of two-electron bound state reads~\cite{review}:
\begin{equation}\label{wcrex}
    \frac{|W_{cr}|}{2t}=\left[1+\frac{2D}{U}\right]^{-1}+(\overline{C}-1)^{-1} \; ,
\end{equation}
where $\overline{C}=1/N\sum_k(1-\gamma_k/z)^{-1}$ is the Watson
integral. This is an exact result. Remembering that $\overline{C}$
is divergent for lattices of dimensions $d=1$ and $d=2$ we can
see, that for $n\rightarrow 0$ $\mu/D\rightarrow -1$,
$I\rightarrow 0$ (Eq.~(\ref{wcrmy})) and Eq.~(\ref{wcrex}) is a
limiting value of the formula Eqs~(\ref{wcrmy}) and
(\ref{wcrmysq}) for $d=1$ and $d=2$, as it should be. Not for
$d=3$ though; this case will be discussed later on.

Let's note that Eqs~(\ref{wcrmy}), (\ref{wcrmysq}) and
(\ref{wcrex}), are valid for any combination of signs of $U$ and
$W$ and for large enough $U<0$ and $W<0$ there is a second branch
of solutions~\cite{friedberg,bak2sol}. The two branches are the
two solutions which realize in the two opposite limits:
$U=+\infty$ and $U=-\infty$ (or $W=\pm\infty$). The formal
equations and their solutions in both these limits are the same,
despite completely different physical situation. Nevertheless
these are the specific cases of two distinct solutions.

\section{Results and discussion}
In view of the Randeria's notice, described in the Introduction,
in the case of s-wave symmetry in two dimensions we can use
condition for existence of bound states as a condition for
existence of superconductivity. In Fig.~1 the boundaries expressed
by Eq.~(\ref{wcrmy}) for different lattice dimensionalities and
electron densities are shown. For parameters $U$, $W$ belonging to
the area above the plotted lines (mostly in the 1st quarter of
coordinate system) two-electron bound states, and what follows
s-wave superconductivity in two dimensions, can not exist. The
curves for $n=0$ in all dimensions are exact results.

\begin{figure}[h]%fig1
\includegraphics[height=6cm]{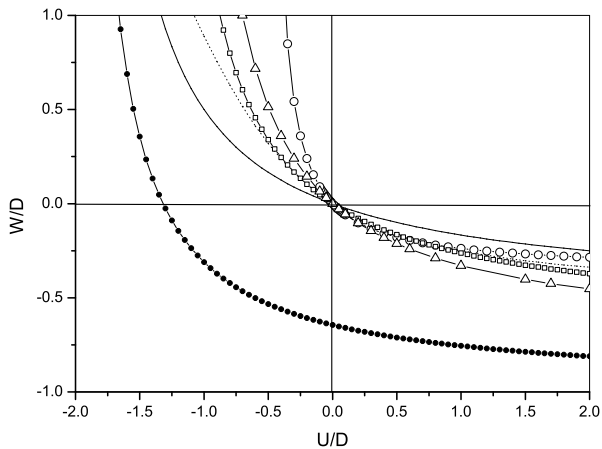}
\raisebox{2.7cm}{\parbox{8cm}{\caption{Critical values for
existence of bound pairs and superconductivity for: $n=0$ for
$d=3$ (black circles - bound pairs only), $n=0$ and rectangular
DOS (full line), $n=0.25$, rectangular DOS and Fock term (dotted
line), $n=0.25$ and rectangular DOS (squares), $n=0.25$ for $d=2$
(triangles) and $n=0.25$ for $d=3$ (white circles -
superconductivity only). Line $n=0$ for $d=2$ is the same as for
the rectangular DOS. Half-bandwidth unit $D=4t$ for rectangular
DOS and for $d=2$ while $D=6t$ for $d=3$.}}}
\end{figure}

As $n$ gets larger, the area of existence of bound pairs increases
for $W>0$ and $U<0$ and decreases in the part of the diagram with
$W<0$ and $U>0$. All curves except the one for $n=0$ in $d=3$ go
through the point with coordinates $(0,0)$. This illustrates the
fact that for $W=0$ infinitesimally small $U$ creates bound state
in $d=2$, while the threshold exists in $d=3$. Nevertheless we do
not have threshold in $d=3$ for $n\neq 0$ -- in agreement with
Randeria's notion about necessity of bound states for
superconductivity only in $d=2$. Let's note that for large $U$ and
$W$ the curves approach the asymptotes -- for curves crossing
through the axes origin the asymptotes are given by the formulas:
$U_{as}/t=-16(n-1)^2/(1+(n-1)^2)$ and $W_{as}/t=-4/(1+(n-1)^2)$.
This is connected with the fact that in the 3rd quarter of the
coordination system for $U<0$ and $W<0$ there exist second
branches of the solutions. As they have higher energy than the
solutions described in Fig.~1 they do not modify the ground state
phase diagram and are not shown here.

The dotted line in Fig.~1 (and in Fig.~2) describes the results of
calculations with inclusion of the Fock term, using
Eq.~(\ref{wcrmysq}) modified by the $(1+pW/zt)$ term, as was
described in the end of Section~2. To simplify the calculations
the Fock term from the normal state was used: $p=n(2-n)$. For
$W>0$ this term broadens the band moving the system more into weak
coupling limit and enlarging the normal state area (opposite
behavior for $W<0$). The same effect can be seen in Fig.~2.

 In
Fig.~2 the boundaries of existence of bound states,
Eq.~(\ref{wcrmysq}) (black symbols), are plotted on a ground state
phase diagram of the extended Hubbard model for $U<0$ and $W>0$,
for arbitrary $n$ and rectangular DOS, together with the phase
boundary PS[CDW/SS]/SS taken from Ref.~\cite{rp} (white symbols).
Above the lines with black symbols s-wave superconductivity can
not exist in $d=2$. This way the superconducting state is
suppressed and normal state (NO) area is introduced into the phase
diagram. Let's note that also the phase separated state PS[CDW/SS]
is "reduced" to the NO phase and not to the CDW phase. This is due
to the fact that the CDW in the PS state is the CDW with $n=1$.
CDW with $n\neq 1$ is unstable, as it has negative
compressibility.

\begin{figure}[h]%fig2
\includegraphics[height=6cm]{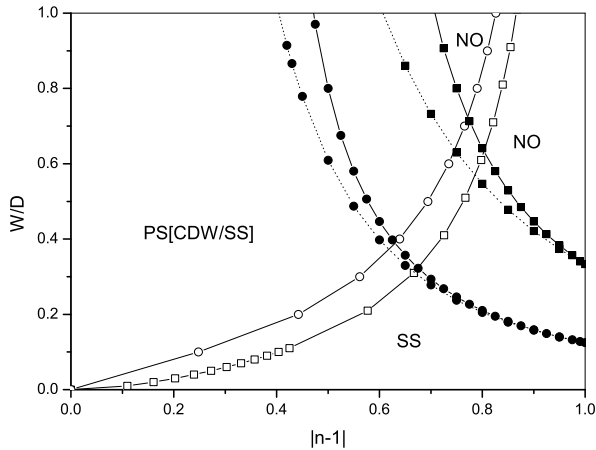}
\raisebox{2.3cm}{\parbox{8cm}{\caption{Phase boundaries for
$U/D=-0.4$ (circles) and $U/D=-0.8$ (squares) calculated for the
rectangular DOS. Black symbols denote boundary of existence of
bound state, white symbols boundary between singlet
superconductivity (SS) and phase separated area PS[CDW/SS] (with
charge density wave CDW and SS). Symbols on dotted lines show the
results of calculations including Fock term.}}}
\end{figure}

Another thing to note is the threshold for appearance of bound
states for $n=0$, which increases with increasing $|U|$. The phase
diagram is modified only for intermediate values of $|U|$ and
$|W|$, smaller from their asymptotic values $|U_{as}|$ and
$|W_{as}|$. For $|U|$ or $|W|$ larger than these values, bound
states exist for arbitrary value of the other parameter, in
agreement with Fig.~1.

The calculations in Ref.~\cite{rp} consider only pure, on-site
s-wave pairing. Including $\Delta_\gamma$ (Eq.~(\ref{delta1}))
into calculations does not change much the described PS[CDW/SS]/SS
boundary -- $\Delta_\gamma$ is two orders of magnitude smaller
than $\Delta_0$ on this boundary. Inclusion of Fock term into the
calculations of the bound states, results in extending the area of
the normal phase, as was mentioned before. This effect increases
with increasing $|n|$ and $W$.

In conclusion it was shown, how the analytical (and exact for
$n=0$) formulas for bound two-electron states can be used for the
modification of the $U<0$, $W>0$ part of the phase diagram of the
extended Hubbard model. The main result of this approach is
suppression of the superconducting and phase separated areas in
favor of the normal phase around half-filled band for intermediate
values of $|U|$ and $|W|$, larger than threshold values and
smaller than $|U_{as}|$ and $|W_{as}|$.
\vspace{-5mm}

\begin{acknowledgements}
I acknowledge discussions with R. Micnas and S. Robaszkiewicz and
support from the Foundation for Polish Science. This work was also
supported by the Polish State Committee for Scientific Research
(KBN), Project No. 1 P03B 084 26.
\end{acknowledgements}

\end{document}